\documentstyle[12pt,psfig,axodraw]{article}

\advance\textheight by \topskip
\newcommand{\newc}{\newcommand}

\newcommand{\be}{\begin{equation}}
\newcommand{\ee}{\end{equation}}
\newcommand{\br}{\begin{eqnarray}}
\newcommand{\er}{\end{eqnarray}}
\newcommand{\ba}{\begin{array}}
\newcommand{\ea}{\end{array}}
\newcommand{\bi}{\begin{itemize}}
\newcommand{\ei}{\end{itemize}}
\newcommand{\bn}{\begin{enumerate}}
\newcommand{\en}{\end{enumerate}}
\newcommand{\bc}{\begin{center}}
\newcommand{\ec}{\end{center}}

\newcommand{\ar}{\rightarrow}

\newcommand{\Dir}{\kern -6.4pt\Big{/}}
\newcommand{\Dirin}{\kern -10.4pt\Big{/}\kern 4.4pt}
\newcommand{\DDir}{\kern -10.6pt\Big{/}}
\newcommand{\DGir}{\kern -6.0pt\Big{/}}
\def\Ecm{\ifmmode{E_{\mathrm{cm}}}\else{$E_{\mathrm{cm}}$}\fi}
\def\gluino{\ifmmode{\mathaccent"7E g}\else{$\mathaccent"7E g$}\fi}
\def\photino{\ifmmode{\mathaccent"7E \gamma}\else{$\mathaccent"7E \gamma$}\fi}
\def\mgluino{\ifmmode{m_{\mathaccent"7E g}}
             \else{$m_{\mathaccent"7E g}$}\fi}
\def\taugluino{\ifmmode{\tau_{\mathaccent"7E g}}
             \else{$\tau_{\mathaccent"7E g}$}\fi}
\def\mphotino{\ifmmode{m_{\mathaccent"7E \gamma}}
             \else{$m_{\mathaccent"7E \gamma}$}\fi}
\def\ML{\ifmmode{{\mathaccent"7E M}_L}
             \else{${\mathaccent"7E M}_L$}\fi}
\def\MR{\ifmmode{{\mathaccent"7E M}_R}
             \else{${\mathaccent"7E M}_R$}\fi}

\def\AJ(#1,#2,#3){Astrophysical.\ Jori. \issue(#1,#2,#3)}

\def\issue(#1,#2,#3){{\bf #1}, #2 (#3)} % AIP format! Vol, page (Year)
\def\PRD(#1,#2,#3){Phys.\ Rev.\ D \issue(#1,#2,#3)}
\def\NPB(#1,#2,#3){Nucl.\ Phys.\ B \issue(#1,#2,#3)}
\def\JP(#1,#2,#3){J.\ Phys.\issue(#1,#2,#3)}
\def\PL(#1,#2,#3){Phys.\ Lett. \issue(#1,#2,#3)}
\def\PLB(#1,#2,#3){Phys.\ Lett.\ B  \issue(#1,#2,#3)}
\def\ZP(#1,#2,#3){Z.\ Phys. \issue(#1,#2,#3)}
\def\ZPC(#1,#2,#3){Z.\ Phys. \ C  \issue(#1,#2,#3)}
\def\PREP(#1,#2,#3){Phys.\ Rep. \issue(#1,#2,#3)}
\def\PRL(#1,#2,#3){Phys.\ Rev.\ Lett. \issue(#1,#2,#3)}
\def\MPL(#1,#2,#3){Mod.\ Phys.\ Lett. \issue(#1,#2,#3)}
\def\RMP(#1,#2,#3){Rev.\ Mod.\ Phys. \issue(#1,#2,#3)}
\def\SJNP(#1,#2,#3){Sov.\ J. \ Nucl.\ Phys. \issue(#1,#2,#3)}
\def\CPC(#1,#2,#3){Comp.\ Phys. \ Comm. \issue(#1,#2,#3)}
\def\IJMPA(#1,#2,#3){Int.\ J. \ Mod. \ Phys.\ A \issue(#1,#2,#3)}
\def\MPLA(#1,#2,#3){Mod.\ Phys.\ Lett.\ A \issue(#1,#2,#3)}
\def\PTP(#1,#2,#3){Prog.\ Theor.\ Phys. \issue(#1,#2,#3)}
\def\RMP(#1,#2,#3){Rev.\ Mod.\ Phys. \issue(#1,#2,#3)}
\def\NIMA(#1,#2,#3){Nucl.\ Instrum.\ Methods \ A \issue(#1,#2,#3)}
\def\JHEP(#1,#2,#3){J.\ High \ Energy \ Phys. \issue(#1,#2,#3)}
\def\EPJC(#1,#2,#3){Eur.\ Phys.\ J. \ C \issue(#1,#2,#3)}
\def\NJP(#1,#2,#3){New.\ J.\ Phys \issue(#1,#2,#3)}

\def\PROP(#1,#2,#3){Prog.\ Part.\ Nucl.\ Phys. \issue(#1,#2,#3)}
\newc{\PRDR}[3]{{Phys. Rev. D} {\bf #1}, Rapid 
Communications, #2 (#3)}
\def\Ord{\lower .7ex\hbox{$\;\stackrel{\textstyle <}{\sim}\;$}}
\def\OOrd{\lower .7ex\hbox{$\;\stackrel{\textstyle >}{\sim}\;$}}
\def \chonepl {{\tilde\chi_1^+}}
\def \chonemi {{\tilde\chi_1^-}}
\def \chplus {{\tilde\chi^+}}
\def \chminus {{\tilde\chi^-}}
\def \ch1pm{{\tilde\chi_1}^{\pm}}

\def \LSP{\tilde\chi_1^0}
\def \LSPsecond{\tilde\chi_2^0}

\def \MSL1{m_{\tilde\l_1}}
\def \MSMU1{m_{\tilde\mu_1}}

\def \LSP{\tilde\chi_1^0}

\def \mhf{m_{1/2}}
\def \MST{m_{\tilde t_1}}

\def\t1{\tilde{t_1}}
 
\newc{\rpv}{{\not\!\!R_p}}
\newc{\rpvm}{{\not R_p}}
\newc{\ttbar}{\mbox{$t\overline{t}$}}
\newc{\qbar}     {\mbox{$\overline{q}$}}
\newc{\squark}   {\mbox{$\tilde{q}$}}
\newc{\sqbar}    {\mbox{$\bar{\tilde{q}}$}}
\newc{\msquark}  {\mbox{$M(\squark)$}}
\newc{\csquarkl} {\mbox{$\tilde{c}_L$}}
\newc{\mcsl}     {\mbox{$M(\csquarkl)$}}
\newc{\ssb}      {\mbox{$\squark\overline{\squark}$}}
\newc{\csquark}  {\mbox{$\tilde{c}$}}
\newcommand{\tsquark}  {\mbox{$\tilde{t}$}}

\newc{\ttbone}   {\mbox{$\tsquark_1\overline{\tsquark}_1$}}
\newc{\chione}   {\mbox{$\tilde{\chi}_{1}^{\pm}$}}
\newc{\mchione}  {\mbox{$M(\tilde{\chi}_{1}^{\pm})$}}
\newc{\mstopo}   {\mbox{$m_{\tilde{t}_1}$}}
\newc{\mcone}    {\mbox{$M(\tilde{\chi}_{1}^{\pm})$}}
\newc{\none}     {\mbox{$\tilde{\chi}_{1}^0$}}
\newc{\mchio}    {\mbox{$M(\none)$}}
\newc{\lsp}      {\mbox{$\tilde{\chi}_{1}^0$}}
\newc{\mz}       {\mbox{$M_0$}}
\newc{\mo}       {\mbox{$M_{1/2}$}}
\newc{\lamp}     {\mbox{$\lambda_{121}'$}}
\newc{\tev}  {\mbox{$\;{\rm TeV}$}}
\newc{\gevc} {\mbox{$\;{\rm GeV}/c$}}
\newc{\gevcc}{\mbox{$\;{\rm GeV}/c^2$}}
\newc{\chis}{\mbox{$\chi^{2}$}}
\newc{\ifb}{\mbox{${\rm fb}^{-1}$}}
\newc{\ipb}{\mbox{${\rm pb}^{-1}$}}
\newc{\met}{\mbox{${E\!\!\!\!/_T}$}}
\newc{\intlum}{\mbox{${ \int {\cal L} \; dt}$}}
\newc{\et}{\mbox{$E_T$}}
\newc{\modulus}[1]{\left| #1 \right|}
\newc{\bp}{\mbox{$b'$}}
\newc{\lxy}{\mbox{$L_{xy}$}}
\newc{\dedx}{\mbox{${\rm d}E/{\rm d}x$}}
\newc{\R}{$R$}
\newc{\charginom}{M_{\tilde \chi}^{+}}
\newc{\mue}{\mu_{\tilde{e}_{iL}}}
\newc{\mud}{\mu_{\tilde{d}_{jL}}}

\newc{\barr}{\begin{eqnarray}}
\newc{\earr}{\end{eqnarray}}
\newc{\beq}{\begin{equation}}
\newc{\eeq}{\end{equation}}
\newc{\ra}{\rightarrow}
\newc{\lra}{\longrightarrow}
\newc{\lam}{\lambda}
\newc{\eps}{\epsilon}
\newc{\gev}{\,GeV}
\newc{\eq}[1]{(\ref{eq:#1})}
\newc{\eqs}[2]{(\ref{eq:#1},\ref{eq:#2})}
\newc{\etal}{{\it et al.}\ }
\newc{\ibid}{{\it ibid}.}
\newc{\eg}{{\it e.g.}\ }
\newc{\ie}{{\it i.e.}\ }
\newc{\nonum}{\nonumber}
\newc{\lab}[1]{\label{eq:#1}}
\newc{\dpr}[2]{({#1}\cdot{#2})}
\newc{\lsimeq}{\stackrel{<}{\sim}}
\newc{\gsimeq}{\stackrel{>}{\sim}}
\newc{\half}{\frac{1}{2}}

\newc{\rpvs}{{\not R_p}}
\newc{\rp}{{${R_p}$}}
\newc{\kap}{\kappa}
\newc{\ptmiss}{/ \hskip-7pt p_T}
\newc{\mgut}{M_U}
\newc{\wt}{\widetilde}
\newc{\gl}{\wt g}
\newc{\mgl}{m_{\gl}}
\newc{\cnone}{\wt\chi^0_1}
\newc{\cpmone}{\wt \chi^{\pm}_1}
\newc{\mcpone}{m_{\cpone}}
\newc{\mcpmone}{m_{\cpmone}}
\newc{\mcnone}{m_{\cnone}}
\newc{\bit}{\begin{itemize}}
\newc{\eit}{\end{itemize}}
\newc{\delgs}{\delta_{GS}}
\newc{\mth}{m_{3/2}}
\newc{\bea}{\begin{eqnarray}}   \newc{\eea}{\end{eqnarray}}
\newc{\baa}{\begin{array}}      \newc{\eaa}{\end{array}}
\newc{\thetaw}{\theta_W}
\newc{\call}{{\cal L}}
\newc{\mplanck}{M_{\rm P}}
\newc{\dmchi}{\Delta m_{\tilde\chi}}
\newc{\cpone}{\wt \chi^+_1}
\newc{\cmone}{\wt \chi^-_1}
\newc{\gam}{\gamma}
\newc{\cntwo}{\wt\chi^0_2}
\newc{\lampp}{\lam^{\prime\prime}}
\newc{\llamp}{\lam^{\prime}}
\newc{\wtil}{\widetilde}
\newc{\glsp}{$\wtil g$-LSP}
\newc{\mev}{~{\rm MeV}}
\newc{\msq}{m_{\squark}}
\newc{\anti}{\overline}
\newc{\mtil}{\widetilde m}
\newc{\mt}{m_t}
\newc{\mw}{m_W}
\newc{\wbp}{W^+}
\newc{\rts}{\sqrt{s}}
\newc{\wm}{W^-}
\newc{\fbi}{~{\rm fb}^{-1}}
\newc{\chitil}{\wt\chi}
\newc{\cmtwo}{\wt \chi^-_2}
\newc{\cnthree}{\wt\chi^0_3}
\newc{\cnfour}{\wt\chi^0_4}
\newc{\pbi}{~{\rm pb}^{-1}}
\newc{\pb}{~{\rm pb}}
\newc{\etmiss}{/ \hskip-7pt E_T}
\newc{\tanb}{\tan\beta}
\newc{\tb}{\tan\beta}
\newc{\mhalf}{m_{1/2}}
\newc{\mcntwo}{m_{\cntwo}}
\newc{\mcpmtwo}{m_{\cpmtwo}}
\newc{\mcnthree}{m_{\cnthree}}
\newc{\mcnfour}{m_{\cnfour}}
\newc{\cpmtwo}{\wt \chi^{\pm}_2}
\newc{\vev}[1]{{\left\langle #1\right\rangle}}
\newc{\mtilq}{m_{\tilde q}}
\newc{\mtill}{m_{\tilde\ell}}
\newc{\cost}{\cos{\theta_{\tilde t}}}
\newc{\lamepd}{\lam'_{131}}
\def\lapp{\mathrel{\rlap{\raise.5ex\hbox{$<$}}
                    {\lower.5ex\hbox{$\sim$}}}}
\def\gapp{\mathrel{\rlap{\raise.5ex\hbox{$>$}}
                    {\lower.5ex\hbox{$\sim$}}}}
\begin{document}
\begin{titlepage}
%\begin{center}
%{\Large DRAFT V1.2}
%\end{center}
\begin{flushright}
%%\today\\
\texttt{hep-ph/0509171} \\
\texttt{CU-PHYSICS-19/2005}\\
\texttt{JU-PHYSICS-09/2005}\\
\end{flushright}
\vskip .6cm
\begin{center}
{\Large\bf  
Top squark and neutralino decays in a R-parity violating model
constrained by neutrino oscillation data
}\\[1.00cm]
{\large Siba Prasad Das$^{a,}$ \footnote{\it spdas@juphys.ernet.in},
Amitava Datta} $^{b,}$ \footnote{\it adatta@juphys.ernet.in},
%\\[0.3 cm]
{and}
%\\[0.3 cm]
{\large Sujoy Poddar} $^{b,}$ \footnote{\it 
sujoy@juphys.ernet.in}
{\\[0.3 cm]}
{\it $^a$Department of Physics, University of Calcutta, Kolkata- 700 009, India}\\[0.1cm]
{\it $^b$Department of Physics, Jadavpur University,
Kolkata- 700 032, India}\\[0.3cm]
\end{center}
\vspace{.1cm}

\begin{abstract}
{\noindent\normalsize
In a R-parity violating (RPV)  model of neutrino mass with three bilinear
couplings $\mu_i$
and three trilinear couplings $\llamp_{i33}$, where $i$ is the lepton index,
we find six generic scenarios each with a distinctive pattern
of the trilinear couplings consistent with the oscillation data. These
patterns may be reflected in direct RPV decays of the lighter top squark
or in the RPV decays of
the lightest superparticle, assumed to be the lightest neutralino.
 Typical signal sizes at the
Tevatron RUN II and the LHC  have been estimated and the results turn out 
to be encouraging.
}
\end{abstract}

\end{titlepage}

\textheight=8.9in

\section*{1.~Introduction}

~~~Neutrino oscillations have  been observed in different experiments \cite{other}, 
from which it is confirmed that the neutrinos have tiny masses, several orders 
of magnitude smaller 
than  any other fermion mass in the Standard 
Model (~SM~). If right handed neutrinos are introduced in the SM one can formally get
Dirac masses of the neutrinos. But the corresponding Yukawa couplings must be    
 unnaturally small.

The SM being unable to naturally explain the origin of very small 
neutrino masses 
,~R-parity violating (RPV) supersymmetry (SUSY) \cite{susy, Barbier} could be 
a viable  alternative.  In the lepton number violating version of these models,
~the neutrino masses (Majorana type) and mixing angles 
\cite{numassold,subhendu} can 
naturally arise  without requiring  the right handed neutrinos. 

In the minimal supersymmetric extension of the SM (MSSM) there is also a R-parity
conserving (RPC) sector with Yukawa like couplings and soft breaking masses/
mass parameters,
which govern the masses of the superpartners of the SM particles 
(the sparticles).
As demanded by the naturalness argument, these masses should be of the order of 
one TeV. The theoretical predictions for the
neutrino masses and mixing angles depend on these RPC parameters,
 in addition to the parameters of the RPV sector.  Thus the exciting program of
 sparticle searches and the reconstruction of their masses 
at the on going ( Tevatron RUN II) and the upcoming (the
Large Hadron Collider (LHC) or the International Linear Collider (ILC))
accelerator experiments have the potential of testing the RPV models of $\nu$- 
mass.

In R-parity conserving( RPC) SUSY the lightest supersymmetric particle (LSP) is 
predicted to be stable and a carrier of missing transverse energy ($\met$).
In contrast the  model under consideration allows the LSP  to decay via RPV 
couplings into
lepton number violating modes. The multiplicity
of particles in any event is, therefore, much larger 
compared to the corresponding event in RPC SUSY containing the stable LSP
leading to distinct collider signatures. Moreover,  since the LSP is not
necessarily a carrier of $\met$ the reconstruction of sparticle masses appears 
to be less problematic than that in the SM.

Another characteristic signature the model under consideration is the
direct decays of sparticles other than the LSP into lepton number violating
channels. Apparently the stringent constraints on the RPV couplings obtained from
the neutrino data (see below) suggest that the branching ratios(BRs) of these
decays will be highly suppressed compared to the competing RPC decays.  One
notable exception is the direct RPV decay of the lighter top squark
\cite{nmass,biswarup,naba,shibu}, if it happens to be the next to lightest
supersymmetric particle(NLSP). This assumption is theoretically well-motivated due
to large mixing effects in the top squark mass matrix. In this case the RPC decays
of the top squark are also suppressed (to be elaborated later) and can
naturally compete with the RPV decays even if the underlying couplings are as
small as that required by the neutrino data. Thus the 
competition among different
decay modes of the lighter top squark, which may be observed during Tevatron 
RUN II, is a hallmark of RPV models of neutrino mass \cite{shibu}.

Of course the see saw mechanism \cite{seesaw} which can be naturally
implemented in any grand unified theory(GUT) not necessarily supersymmetric,
provides an elegant alternative explanation of small neutrino masses. Unfortunately the
simplest version of this theory, a GUT with a grand desert, predicts a low
energy spectrum practically identical with that of the SM.  
Thus there is no testable prediction outside the neutrino sector.

In view of the large number of RPV parameters, constraints on them 
\cite{Barbier,allanach} are usually obtained from the experimental data by
employing a simplifying assumption. It is assumed that only a minimal set of
the parameters contributing to the observables under study are numerically
significant. Following this approach one usually analyzes some 
benchmark scenarios, each consisting of a minimal set of RPV 
parameters at the weak scale \cite{abada,gautam},  capable
of reproducing a phenomenologically viable neutrino mass matrix. Constraints
on the parameters belonging to each scenario are then obtained by using
the neutrino oscillation data \cite{abada,gautam}.

Among the examples in Ref. \cite{abada},~we have focused on 
a specific model
with three trilinear couplings $\lambda'_{i33}$ (~where $i=$ 1,2,3 is 
the lepton generation index~) and three bilinear RPV parameters $\mu_{i}$.The  
stringent upper bounds on the trilinear (bilinear) couplings are
$\sim 10^{-4}$ (~$\sim 10^{-4}GeV$~). As a result the contributions of these
couplings to most low energy processes except LSP decays are negligible. As
already mentioned, a notable exception could be the direct RPV decay of the
lighter top squark \cite{naba,shibu} into a $b$-quark  and a charged lepton.
Moreover, by reconstructing the lepton - jet invariant mass the lepton number
violating nature of the decay can be directly established \cite{shibu}. A model
independent estimate of the minimum observable branching ratio (MOBR) of the 
channel $\t1
\ar e^+\bar{d}$ as a function of the lighter top squark mass ($\mstopo$) for
Tevatron RUN II was also presented \cite{shibu}. This estimate was then
translated into the magnitudes of the $\llamp$s for representative
values of the parameters in the RPC sector and they were found to be
close to the bounds obtained from the oscillation data.

In this paper we go a step beyond simple estimates and obtain testable
quantitative predictions. We find various  combinations of three 
$\llamp$-type couplings allowed by the neutrino data. We do this by
randomly generating $10^9$ sets of the above six RPV couplings for
representative values of the parameters in the RPC sector and identify the sets
consistent with the oscillation data \cite{global}. Our numerical results are 
checked by
analytical calculations in a simple approximation (see sections 2 and 3 for the
details). An interesting common feature of all the allowed sets is that one
$\mu_i$ is much smaller than the other two and for 
each such scenario there are 
two characteristic patterns of the 
$\lambda'_{i33}$ couplings. We thus have six
generic scenarios. In each scenario the 
relative BRs of the three decay modes 
$\t1 \ar l^+_j \bar{b}$ reflect the underlying model. 
Thus if some of the  RPV decay channels of $\t1$ are observed with BRs as
predicted, a particular generic scenario can be vindicated. Of course more
definite conclusion can be drawn if complementary information about 
the RPC sector, 
the masses of the sparticles in particular,
can be obtained kinematically 
(by measuring invariant masses of the final state particles, edges
of the energy distributions of some of the  decay products etc). Some simple examples 
will be given later.

In our numerical work, we have chosen the parameters of the RPC sector
corresponding to several scenarios classified according to the properties of the
electroweak gaugino sector which determine the tree level neutrino masses ( see
section 4 for the details). In each case the lightest neutralino is assumed to be
the LSP. We then study the predicted competition among various RPC 
\cite{hikasa,boehm} and RPV decay modes of the top squark NLSP for 
RPV couplings allowed by the 
neutrino data. We find that many of the solutions predict BRs larger 
than the estimated MOBR at Tevatron in \cite{shibu}. We have also 
calculated the number of some typical signal events, where the expected  
  background is low from top squark production and 
decay at Tevatron RUN II and LHC.  

If on the other hand no signal is seen during Tevatron RUN II the 
allowed parameter space(APS)  
will be significantly squeezed. From the limits on the trilinear couplings
obtained from Tevatron RUN I and RUN II \cite{naba} and from the projected
sensitivity of RUN II data to these couplings \cite{shibu}, it seems that
$\lambda'_{i33} \sim 10^{-4}$ can be ruled out for lighter top squark masses
within the kinematic reach of the Tevatron.

 If the top squark is not the NLSP and the RPV couplings are as small as that
required by the current neutrino data \cite{global}, the decay of the LSP 
would be
the only signature of the RPV model of neutrino mass. However, the amount of
information that may be extracted will depend on parameters of the RPC sector, in
particular on the LSP mass. If the LSP is lighter than the top quark, then it
decays via the modes

\be
\LSP  \ar \nu_{\l} b \bar b  
\label{lsp3bdk}
\ee
where $l$ = e , $\mu$ or $\tau$. 
Since the neutrinos are not detectable the branching ratios of the individual
channels can not be measured. The decay length of the LSP depends on many 
model parameters belonging to the RPC and RPV sectors 
and pin pointing the underlying model of neutrino mass  from one 
observable may not be an easy task (see, however, \cite{porod}).
In addition the lepton number violating nature of the underlying
interaction can not be directly established  since the neutrinos escape the 
detector. 

The signatures will be unambiguous if the $\LSP$ happens to be heavier than the
top quark. In this case apart from the decay channel in Eq. (1) the LSP will also
decay into a charged lepton ( e, $\mu$ or $\tau$ ), the top quark and the
anti-bottom quark, a clearly lepton number violating signature. This 
observation motivates us to  also scan the parameter space where the LSP is 
heavier than the top quark.  
Of course the decay modes involving the top quark
will be phase space suppressed compared to the ones in Eq. (1).  
Nevertheless our
computations show that the branching ratios of the three modes involving the
t-quark are numerically significant over the entire parameter space allowed by 
the
oscillation constraints. Since the LSP is rather heavy in this scenario it is
unlikely to be produced at the Tevatron. However, observation of all four
modes and measurement of their branching ratios at the LHC or ILC 
will provide crucial
tests of the underlying model of neutrino mass. We have calculated the size of
some typical signals from the  production of 
electroweak gaugino pairs at the LHC.

The plan of the paper is as follows. In section 2 we establish the 
notations and briefly review the
neutrino mass matrix in the RPV model under study. In section 3 
we identify the six generic scenarios of the RPV sector compatible with
the neutrino data and 
using several representative values of the parameters of the RPC sector 
obtain sets of RPV parameters allowed by the oscillation data.
In section 4 the top squark decays are studied in different scenarios.
The LSP decays are analyzed in section 5. Our conclusions and future outlooks 
are summarized in section 6.

\section*{2.~Neutrino mass matrix at the tree and loop level}

R-parity is a multiplicative quantum number defined as follows \cite{Barbier},

\beq
R_p=(-1)^{3B+L+2S},
\eeq
where $B$ is the baryon number, $L$ is the lepton-number and $S$
denotes the spin. For particles $R_p$ = +1 and for sparticles $R_p$ = -1.

~In general the superpotential of the MSSM may contain RPV terms which violate
both B and L conservation.~ This leads to catastrophic proton decay with a mean
life time not allowed experimentally.~ All B and L violating terms can be
removed from the superpotential by imposing R-parity as a symmetry.~ The
resulting model is known as the R-parity conserving  MSSM.
~~In order to prevent proton decay it is,~however,~sufficient to remove either
B-violating or L-violating terms by imposing appropriate discrete symmetries.
Models with B-violating terms only can not generate neutrino masses.
 As discussed in the introduction we have focused on a specific model of 
neutrino mass with $\llamp$ type couplings only. 
~ The general R-parity violating superpotential of our interest 
takes the form:

\barr
W&=&W_{MSSM}+W_\rpvs, \\
W_\rpvs&=& 
\lam'_{ijk} \eps_{ab}
L_i^aQ_j^b{D}_k^c +\mu_i \eps_{ab} L_i^a H_2^b.  \lab{superpot}
\earr
Here,~ $W_{MSSM}$\cite{susy} is the usual superpotential of the MSSM containing the
terms which give mass to the SM fermions.~
 The  $i,j,k=1,2,3$ are generation indices and
 $a,b=1,2$ are $SU(2)$ indices and 'c' denotes charge conjugation. The
$\lam'$s are dimensionless trilinear RPV  Yukawa like couplings, $\mu_i$'s
 are bilinear RPV terms \cite{Barbier,chang feng} 
with dimensions of mass, which determines the amount of 
mixing between the lepton and Higgs superfields.~In Eq.(4) $L$,~$Q$~and $H_2$ 
denote,~respectively
,~$SU(2)_L$ doublet lepton,~quark and up type higgs superfields and $D$ 
is the $SU(2)_L$ singlet down type quark superfields. One can also 
construct models 
of $\nu$-mass with RPV $\lam$-type couplings \cite{Barbier,numassold,
subhendu,REF2} only. The phenomenology of such models will be quite distinct 
from that discussed in this paper.

 The tree level and loop level neutrino mass matrices are given below
\cite{Barbier,subhendu,abada}:

\br\hspace*{-0.5cm}
{\cal{M}}^{\mathrm{tree}}_{\nu_{ij}}= C  \mu_i  \mu_j ,
 \label{MRptree}
\er
and $C$ is given by :
\begin{equation}
C = g_{2}^{2}{(M_{1} +
 \tan^2\theta_{W} M_{2})\over 4
\det M}\ v_{d}^{2}
\end{equation}
where $g_2$ is the $SU(2)$ gauge coupling, $M_1, M_2$ are the $U(1)$ and
$SU(2)$ gaugino masses respectively and $\det M$ is the determinant
of the R-parity conserving neutralino mass matrix. Here we are working in the
basis where the sneutrino vacuum
expectation values (vevs) are  zero, $v =\sqrt{v_d^2 + v_u^2}$,
~where $v_d$ ($v_u$) is the vev of  down (up) type Higgs field.
~The one loop mass matrix is given by :

 \br\hspace*{-0.5cm}
 {\cal{M}}^{\mathrm{loop}}_{\nu}= \left(
  \begin{array}{lll}
   K_2\lambda'^2_{133}&
   K_2\lambda'_{133}\lambda'_{233} &
    K_2\lambda'_{133}\lambda'_{333}\\
     K_2\lambda'_{133}\lambda'_{233}&
      K_2\lambda'^2_{233}&
       K_2\lambda'_{233}\lambda'_{333} \\
       K_2\lambda'_{133}\lambda'_{333} &
   K_2\lambda'_{233}\lambda'_{333}
       & K_2\lambda'^2_{333}
          \end{array}
            \right) ,
            \label{MRploop}
             \er

where $K_2$ is given by :

\begin{eqnarray}
K_2&=&3  {X_b\over 16 \pi^2} {f(x_q) \over M_{q2}^{(3)^2}} \left(  m_{b}^2 \right)\ .
\label{AAB}
\end{eqnarray}
with
\bea
f(x)=-{\ln \ x\over 1-x},\quad
%x_\ell^{(p)}= \left({M_1^{(p)}\over M_2^{(p)}}\right)^2\quad, \quad
x_q= \left({M_{q1}^{(3)}\over M_{q2}^{(3)}}\right)^2  {\mathrm{and}}\quad 
\eea
\bea
X_{b}= A_{b}- \mu \tan\beta .
\eea

In the above   $X_{b}$ is the off-diagonal mixing  term in the 
b-squark mass matrix  
, $A_b$  is the soft trilinear term for the bottom squarks, $\tan\beta$ is the
ratio of $v_u$ to $v_d$, $\mu$ is the
higgsino mass parameter
,  $m_b$~ is mass of the bottom quark 
. $M_{q1,q2}^{(3)}$ are the masses of the two b - squark mass eigenstates.

~In this work  we shall assume that  the masses of the right handed and left 
handed squarks are equal. 
The same assumption applies to the slepton sector.

~~ Thus,~ we take  mass matrix up to the one-loop to be
  \bea
  {\cal{M}_{\nu}}={\cal{M}}^{\mathrm{tree}}_{\nu}
+{\cal{M}}^{\mathrm{loop}}_{\nu}.
\label{Moneloop}
\eea

It has been noted in the literature that there may be other loop contributions
to the neutrino mass matrix \cite{loop,subhendu}. For example, 
the soft breaking RPV
bilinear terms may contribute to some of the loops \cite{subhendu}. As has 
already been mentioned the large number of free parameters compels one to
work in  benchmark scenarios with a limited number of RPV parameters.    

In Table [1] we present the neutrino data that has been used for the numerical 
work in this paper \cite{global}. We shall consider the data at the 2$\sigma$ 
level.   
The notations used are as follows \cite{angl}: the neutrino
mass squared differences are $\Delta^2_{sol}= \Delta m^2_{21}=
 |{m^2_2-m^2_1}|$ and
$\Delta m^2_{atm}=\Delta m^2_{31}=|{m^2_3-m^2_1}|$ respectively,~where $m_1$,~$m_2$
and $m_3$ are the three  eigenvalues of the neutrino mass matrix in Eq. 1 
\cite{abada}. The mixing angles are extracted from the eigenvectors 
corresponding to appropriate mass eigenvalues.

\begin{table}[!htb] 
\begin{center}

\begin{tabular}{|c|c|c|c|c|}
       \hline
        parameter & best fit & 2$\sigma$ & 3$\sigma$ & 4$\sigma$
        \\
        \hline
         $\Delta m^2_{sol}[10^{-5} eV^2]$ & 8.1 & 7.5--8.7 & 7.2--9.1 & 7.0--9.4\\
         $\Delta m^2_{atm}[10^{-5} eV^2]$ & 2.2 & 1.7--2.9 & 1.4--3.3 & 1.1--3.7\\
         $\sin^2\theta_{12}$ & 0.30 & 0.25--0.34 & 0.23--0.38 & 0.21--0.41\\
         $\sin^2\theta_{23}$ &0.50 & 0.38--0.64 & 0.34--0.68 & 0.30--0.72 \\
         $\sin^2\theta_{13}$ &0.000 &  $\leq$ 0.028 & $\leq$ 0.047  & $\leq$ 0.068 \\
        \hline
\end{tabular}
\end{center}
   \caption{ Best-fit values,~ 2$\sigma$,~ 3$\sigma$,~and 4$\sigma$ intervals 
for the three
flavour neutrino oscillation parameters from global data analysis \cite{global}
including solar, 
atmospheric, reactor (KamLAND and CHOOZ) and accelerator (~K2K~) experiments 
\cite{other}.}
\end{table}

\section*{3.The six generic scenarios allowed by neutrino data} 

The neutrino mass matrix in section 2 can be recasted 
in the following form: 

\br\hspace*{-0.5cm}
 {\cal{M}}^{\mathrm}_{\nu}= \left(
  \begin{array}{lll}
   D_1 &
   T_1 &
    T_2\\
     T_1&
      D_2&
       T_3 \\
       T_2 &
       T_3
       & D_3
          \end{array}
            \right) ,
            \label{MRptotal}
             \er
  where $T_i$  and $D_i$  are given below :

\bea
  D_1= C \mu_{i}^2 + K_{2} \lambda'^2_{133}\\
  D_2= C \mu_{i}^2 + K_{2} \lambda'^2_{233}\\
  D_3= C \mu_{i}^2 + K_{2} \lambda'^2_{333}\\
  T_1= C \mu_{1} \mu_{2}  + K_{2} \lambda'_{133} \lambda'_{233} \\
  T_2= C \mu_{1} \mu_{3}  + K_{2} \lambda'_{133} \lambda'_{333}\\
  T_3= C \mu_{2} \mu_{3}  + K_{2} \lambda'_{233} \lambda'_{333}
\eea
The oscillation parameters can be easily calculated analytically if any one of the 
$T_i$s  vanish and $D_i \neq 0 $ for all i. 
Thus we tentatively propose the following  generic scenarios :
\\
(a) $\mu_1=0$~ and either $\llamp_{233}=0$ ($T_1=0$) or $\llamp_{333}=0$ 
($T_2=0$)\\
(b) $\mu_2=0$~ and either $\llamp_{133}=0$ ($T_1=0$) or $\llamp_{333}=0$ 
($T_3=0$)\\
(c) $\mu_3=0$~ and either $\llamp_{133}=0$ ($T_2=0$)or $\llamp_{233}=0$ 
($T_3=0$)\\ 

~The above patterns help us to classify different regions of the RPV parameter 
space consistent with the oscillation data in Table [1].

~For illustration we consider a particular hierarchy  $\mu_1 \ll \mu_2$ or~$\mu_3 $ 
(generic scenario (a) ) and  the following representative choice of RPC parameters :
 $M_1$=110, $M_2$=200, $\mu$ =400, $\tan \beta$ =5, $M_{\tilde q}$=400, 
$A_{\tilde  b }$=1000
where all masses and mass parameters are in GeV.
We randomly vary all six RPV parameters within the ranges shown in Table [2]
columns 2 and 3 and generate $10^9$ sets of parameters. Only $\mu_1$ is 
constrained to be rather small. 

\begin{table}
\begin{center}
\begin{tabular}{|c|c|c|c|c|}
    \hline
     & \multicolumn{2}{c|}{Given ranges}&\multicolumn{2}{c|}{Allowed ranges} \\
\cline{2-5} 
RPV parameters   & Max & Min & Max & Min \\
     \hline
      $\mu_1[10^4 eV]$           & 50.0  & 0.001 & 19.0  & 0.10\\
      $\mu_2[10^4 eV]$           & 300.0 & 1.0   & 134.0 & 90.0\\
      $\mu_3[10^4 eV]$           & 300.0 & 1.0   & 178.0 & 89.0\\
      $\lambda'_{133}[10^{-5} ]$ & 12.0  & 0.001 & 4.32  & 3.2\\
      $\lambda'_{233}[10^{-5} ]$ & 12.0  & 0.001 & 9.52  & 0.01 \\
      $\lambda'_{333}[10^{-5} ]$ & 12.0  & 0.001 & 9.67  & 0.002 \\
     \hline

\end{tabular}
\end{center}
\caption {Allowed ranges of RPV parameters in scenario (a)}
\end{table}

We then pick up one solution corresponding to $\mu_1=0.08 \times10^4 eV$,~
$\mu_2 = 140 \times 10^4 eV$,~ $\mu_3=111\times 10^4 eV$,~
$\llamp_{133}=3.94\times10^{-5}$,~$\llamp_{233}=.015 \times10^{-5}$ and
$\llamp_{333}=7.03 \times10^{-5}$. The numerically calculated oscillation
parameters are presented in the second column of Table [3]. In column 3 of
the same table we present the analytically calculated results with
the approximation $\mu_1=0$,~$\llamp_{233}=0$, while other RPV parameters 
are as
above. The agreement between the two sets provides a test of the
reliability of the numerical procedure. Changing the range of variation of 
$\mu_1$ does not lead to any qualitatively new solution with $\mu_1$ 
comparable to $\mu_2$ or $\mu_3$. Thus classifying a
 generic scenario by a
a small magnitude of $\mu_1$ is indeed valid.  

 \begin{table}[!htb]
\begin{center}

\begin{tabular}{|c|c|c|}
  \hline
    Neutrino oscillation  & Numerical results &  Analytical results \\
    parameters & & \\

    \hline

     $\Delta m^2_{sol}[10^{-5} eV^2]$ & 8.65 & 8.80 \\

     $\Delta m^2_{atm}[10^{-3} eV^2]$ &1.74  & 1.69 \\

     $\sin^2\theta_{12}$ & 0.329 & 0.327\\
     
     $\sin^2\theta_{23}$ & 0.437 &0.435 \\
     
      $\sin^2\theta_{13}$ & 0.016  &0.016 \\
      \hline
\end{tabular} 
\end{center} 
\caption{Comparison of Numerical and Analytical results}
\end{table}

 Out of $10^9$ sets of 
randomly generated RPV parameters only  a few satisfy the data
in Table [1]. This illustrates the highly restrictive power of 
the currently available oscillation data 
inspite of the  relatively large errors. 
The allowed ranges of the six RPV parameters are 
presented in  Table [2] columns 4 and 5.
~~ Examining the entire APS corresponding to the  
hierarchy  $\mu_1 \ll \mu_2,~\mu_3 $, where $\mu_2$,~$\mu_3 $ can 
be comparable, one can identify two subclasses under the generic scenario (a). 
The trilinear RPV couplings in the two subclasses are found to satisfy the 
following patterns:\\
\\
($a_1$) $\llamp_{333} >  \llamp_{133} \geq \llamp_{233}$, \\
($a_2$) $\llamp_{233}> \llamp_{133}  \geq \llamp_{333}$.\\
 
In fact  most of the APS $a_1$ ($a_2$ ) 
corresponds to  $\llamp_{233} \ll  \llamp_{133}$ 
  ( $\llamp_{333} \ll  \llamp_{133}$). 
However, there are  exceptions albeit for  relatively small regions of the 
APS, where the two smaller couplings could be of comparable magnitude.

For 
$\mu_2 \ll \mu_1$ or $\mu_3$ (scenario b) and  $\mu_3 \ll 
\mu_1$ or $\mu_2$ (scenario c) each region of the APS is also 
characterized by a specific  pattern of the trilinear couplings 
due to the  constraints imposed by the neutrino data.
If some direct RPV decay modes are observed, these patterns would be 
reflected in the measured  BRs revealing the model  underlying 
neutrino oscillations.\\

For $\mu_2 \ll \mu_1$ or $\mu_3$, the trilinear couplings follow the 
patterns \\
\\
($b_1$) $\llamp_{133} \approx  \llamp_{233} >>  \llamp_{333}$, \\
($b_2$) $\llamp_{233} >  \llamp_{333} >>  \llamp_{133}$. \\

For $\mu_3 \ll \mu_1$ or $\mu_2$, on the other hand, we have \\
\\
($c_1$) $\llamp_{333} >  \llamp_{133} >>  \llamp_{233}$, \\
($c_2$) $\llamp_{333} >  \llamp_{233} >>  \llamp_{133}$. \\

It can be readily checked analytically that even in the most general case when
none of the RPV parameters vanish one eigenvalue is still
zero. Thus analytical solutions are still possible.  
The formulae for the masses and the 
mixing angles are somewhat cumbersome. For scanning the parameter space we 
therefore prefer the numerical method.

It should be noted that there is no straightforward way of determining the 
$\mu_i$ parameters directly from collider 
signals. Thus it is gratifying to note that the scenarios ($a_1$) - 
($c_2$)
can be identified by the decay branching ratios alone provided RPV
decays into charged leptons are observed.

On the face of  it the scenarios ($a_1$) and ($c_1$) look similar. But 
scanning the entire APS in both cases we have found  that\\ 

2.6 \%  $< BR(e) < $ 4.2 \% and 19.7 \% $< BR(\tau)< $ 10.2 \% in 
($a_1$),\\

\noindent 
while\\ 

9.5 \%  $< BR(e) < $ 12.8 \% and 18.4 \% $< BR(\tau)< $ 22.0 \% in 
($c_1$),\\

\noindent
where $BR(e)$ ($BR(\tau)$) refer to the BR of any direct RPV decay mode into
a final state with e ($\tau$). Thus each scenario will have its 
characteristic decay pattern.

\section*{4.The lighter top squark decay} 

~We now study the collider signals that may be triggered by the sets 
of ~$\lambda'$\ ~couplings 
consistent with the neutrino data\cite{global}.
 The signals can be classified into a few patterns 
corresponding to the six generic scenarios  discussed in section 3
and the hierarchy of trilinear couplings associated with them.  
All allowed sets  would lead to the         
RPV decays of the lighter top  squark~$(\tilde t_1)$ in Eq. (19)
 with appreciable branching ratios if it happens to be the NLSP,
which is the case  over a large region of the RPC parameter space.

\be  
(a)~~\tilde t_1 \ra l_i^+ b 
\ee     
where i = 1 - 3. From section 3 it is clear  that 
the relative BRs  of the three leptonic modes  will be different in 
generic scenarios a1) - c2). Hence 
identification of  the RPV parameters 
underlying the model of neutrino mass 
in future experiments is a distinct possibility.
         
~The RPC decay modes of the lighter top- squark       
$(\tilde t_1)$ are listed  below :
\be      
(b)~~\tilde t_1 \ar b \tilde\chi_1^+              
\label{2dk}
\ee
\be
(c)~~\tilde t_1 \ar b \ell \tilde\nu ,~ b \tilde\ell \nu ,~ bW \LSP
\label{3bdk}
\ee
\begin{equation}
(d)~~\tilde t_1 \ar c \tilde\chi_1^0
\label{loopdk}
\end{equation}
\begin{equation}
(e)~~\tilde t_1 \ar b \tilde\chi_1^0 f \bar f'
\label{4bdk}
\end{equation}
 $f$ and $\bar f'$  being a quark-antiquark or $l$-$\bar\nu_l$ pair.
If the lighter top squark is the NLSP, only
the decay modes d) and e) and the last mode of c) are allowed. The 
mode in c) will be phase space suppressed for $\t1$ masses within
the kinematic reach of the Tevatron, which are the main subject of this 
study. Moreover, it will be highly suppressed if the LSP happens to be 
bino like, which is quite natural in popular models like the ones  
with a unified  gaugino mass. We shall not consider this mode in this paper.
The channels in d) and e) have naturally  suppressed widths 
and can very well compete with each other~\cite{boehm} or
with the RPV mode, especially if $\lambda'_{i33}$  is 
$\sim 10^{-4} - 10  ^{-5}$ as required by neutrino data \cite{naba,shibu}.

Our choices of the RPC parameters are guided by the 
gaugino sector.
It is clear from   Eq.(5) section 2 that the parameter C  sets the scale
of  the tree level neutrino mass matrix. This C depends
solely on the parameters
of the gaugino mass matrices. Accordingly we
have chosen the following scenarios.

\begin{enumerate}
\item 

Models in which the lighter chargino ($\tilde\chi_1^{\pm}$)  and the two
lighter neutralinos ($\tilde\chi_1^0$ and $ \tilde\chi_2^0$)  are higgsino
like ($M_1$,$M_2$ $\gg$ $\mu$) and all have approximately the same mass (
$\approx \mu$). Thus it is difficult to accommodate the top squark NLSP
without fine adjustments of the parameters. Thus the LSP decay seems
to  be the only viable collider signature which will be discussed in 
the next section.

\item 
Models in which $\tilde\chi_1^{\pm}$, $\tilde\chi_1^0$ and 
$\tilde\chi_2^0$ 
are gaugino like  ($M_1 < M_2$ $\ll$ $\mu$) 
and the top squark is the NLSP. 

\item  

Models in which $\tilde\chi_1^{\pm}$ and $\tilde\chi_2^0$ are mixed
($M_1$$ < M_2$$\approx$ $\mu$) and the top squark is the NLSP.

\end{enumerate}

In all  models the parameters are so chosen that the lightest 
neutralino ($\LSP$) happens to be the LSP.
Further the squarks belonging to  the right and left  
sectors of all flavours  are assumed to be mass degenerate.
In fact although  the common squark mass, $\mu$ and 
tan $\beta$ occur in both the neutrino and the top squark sector, one
can choose the soft trilinear parameter $A_t$, which does not affect the 
neutrino sector,  to satisfy the top squark NLSP criterion in most cases.
However, attention must be paid so that large values of $A_t$ do not 
lead to a charge color breaking (~CCB~) vacuum \cite{raby}. The BRs of top 
the squark decay modes 
and competition among the RPV and RPC decays would be highly 
indicative of the underlying model.

For model 2, the gaugino like model, 
we use the following representative values of the parameters
of the RPC sector: $M_1=110$, $M_2=200$, $\mu=400$, $\tan \beta =5$,
$M_{\tilde q}=400 $, $A_b=1000$, $A_t=970 $ and $M_{A}=300$ where all
masses and mass parameters are in GeV and $M_{A}$ is the CP odd higgs
mass. The first six parameters, which are the same as the ones used
 in section 3 for numerical illustration, along with the RPV parameters
determine neutrino masses and mixing
angles( see section 2). The last two parameters are required to realize
the top squark
NLSP condition and the CCB condition respectively. It should be 
noted that the BR of the loop decay increases significantly for larger
$M_A$.

Top squark
NLSPs having different masses are realized varying $A_t$.
For this set of RPC parameters the NLSP and the CCB conditions are
satisfied for $940< A_t < 980$. In addition we have used a common
mass of L and R type sleptons 
$M_{\tilde l} =350$ and  $A_{\tau}=1000$ for the computation of the 
BRs of the RPC modes. 

We then randomly generate $10^8$ sets of the six RPV parameters 
in scenarios ( $a_1$ and $a_2$ ) and filter
out the ones allowed by the data in Table [1]. As discussed in section 3 in
all allowed sets there is one dominant coupling ( $\llamp_{233}$ or
$\llamp_{333}$). This
hierarchy will obviously be reflected in the observed BRs.
In Fig. 1 we present the BR of each of the three 
competing decay 
modes in Eq.(19) (including all leptonic channels), Eq.(22) and Eq.(23) 
( including all possible $f$ and $\bar f'$ combinations) and the 
corresponding 
number of allowed solutions out of 
10$^8$ randomly generated parameter sets. This figure clearly     
illustrates the competition among the
three decay modes of the top squark NLSP. Here $A_t =970 GeV$ and 
$m_{\tilde{t}_1}= 181.5 GeV$. The RPV 
decay modes do not dominate although the combined BR of the three RPV 
modes is appreciable ( 20 \% - 30 \%) over most of the APS. It also
follows from this figure that the number of solutions allowed by the $\nu$-
data is indeed a tiny fraction of the total number of 
generated parameter sets. Thus
the available oscillation data is already very resistive in spite of the large
errors.

In Ref.\cite{shibu} top squark pair production followed by their RPV decays 
into $e^{+} e^{-}$ or $\mu^{+} \mu^{-}$ channels were studied for Tevatron 
RUN II. The 
minimum observable BR (MOBR) at Tevatron for the decays
into the e or $\mu$ channel have been estimated as a function of 
$m_{\tilde t_1}$. 
For $m_{\tilde t_1}= 181.5 GeV$ the MOBR  of the ($\tilde t_1 \ra e^{+} 
b $) or ($\tilde t_1 \ra \mu^{+} b $) channel is approximately 20 
\% .

It may be recalled that the analysis of \cite{shibu} was conservative
since only the leading order top squark pair production cross section 
and a total integrated luminosity of only 2000 $pb^{-1}$ were used.
The next to leading order cross section is about 30\% larger and 
accordingly a smaller estimate of the MOBR is expected. The total 
integrated luminosity
during RUN II may be as large as 9000 $pb^{-1}$ \cite{lumi} which can
further lower the MOBR. An
improvement in the observability of RPV decays is also 
expected if ee, $\mu \mu$ and 
e$\mu$ channels are simultaneously analyzed.

For the solutions in 
which decays into the $\tau$ lepton dominate, the combined BR of the
modes involving lighter leptons are below the MOBR. 
Since this happens in a large region of the APS, it will be  worthwhile to
estimate  the MOBR of this channel.
On 
the other hand if the RPV decay into e or $\mu$ dominates, many allowed 
solutions have  BRs close to the MOBR estimated in \cite{shibu}. 

For numerical  illustrations 
we have considered the following points in the APS \\ 

\noindent 
A) $\llamp_{133}=3.72 \times 10^{-5}$,~$\llamp_{233}=3.3 
\times 10^{-5}$ and 
$\llamp_{333}=8.5 \times 10^{-5}$ ( scenario ($a_1$), $\llamp_{133} \approx 
\llamp_{233}$ )\\ 
B)  $\llamp_{133}=4.2 
\times 10^{-5}$,~$\llamp_{233}=1.3 \times 10^{-5}$ and
$\llamp_{333}=8.3 \times 10^{-5}$ ( scenario ($a_1$), $\llamp_{133} > 
\llamp_{233}$ )\\
C) $\llamp_{133}=4.2 
\times 10^{-5}$,~$\llamp_{233}=8.6 \times 10^{-5}$ and
$\llamp_{333}=1.8 \times 10^{-5}$ ( scenario ($a_2$), $\llamp_{233} > 
\llamp_{333}$ ) \\

The BRs of top squark decay modes into different leptonic channels
for the parameter sets A) - C) are given in Table [4].

\begin{table}[!htb]
\begin{center}

\begin{tabular}{|c|c|c|c|c|c|}
       \hline
        Decay modes &$\tilde t_1  \ar  e^+  b$ & $\tilde t_1  \ar  
\mu^+ b$  & $\tilde t_1   \ar  \tau^+   b$  & loop & 4 body     \\
        \hline
BR(A)          &0.03  &0.024  &0.158  &0.69 &0.10 \\
BR(B)          &0.04  &0.004  &0.153  &0.70 &0.11 \\
BR(C)          &0.039  &0.163  &0.007  &0.69 &0.10 \\

        \hline
\end{tabular}
\end{center}
   \caption{BRs of the competing decay modes of the lighter top squark in
the gaugino dominated model ( Model 2, see text for the choice of parameters )}

\end{table}

In Table [5] we present the number of events corresponding to different
dileptonic final states at the Tevatron arising from RPV decays of both 
the produced top squarks. 
 The  number of events are computed for the BRs in Table [4], $\sqrt{s} = 
2$ TeV 
 and an integrated luminosity of 9000 $pb^{-1}$. For
$m_{\tilde t_1}$ = 181.5 GeV the 
 production cross section $\sigma ( p \bar p \ar  \tilde t_1 \tilde t_1^{*})$ 
is  0.41 $pb$ as computed by {\tt CalcHEP v2.1} \cite{CALCHEPcompc4}.
 The signal
\beq
 \tilde t_1   \tilde t_1^* \ar l_i^{+} \bar b l_j^{-}  b
\eeq
is abbreviated as $L_{ij}$, where 
i,j = 1,2,3 for e, $\mu$ and $\tau$ respectively. For i $\neq$ j we 
considered dileptons of all charge combinations.

 \begin{table}[!htb]
\begin{center}

\begin{tabular}{|c|c|c|c|c|c|c|}
       \hline
          &$L_{11}$ & $L_{12}$ & $L_{13}$ &$L_{22}$& $L_{23}$ &$L_{33}$       \\

        \hline
 $\#$ of events (A) &3 &5 &34 &2 & 27 &91 \\
        \hline
 $\#$ of events (B) &5 &1 &44 &0 & 4 &86 \\
        \hline
 $\#$ of events (C) &5 &46 &2 &97 & 8 & 0\\
      
        \hline
\end{tabular}
\end{center}
\caption{Typical sizes of opposite sign dileptons of different
flavour combinations at the Tevatron 
from top squark pair production using the BRs in Table [4].}
\end {table}

It is expected that the backgrounds can be 
suppressed to the desired level
by employing standard kinematical cuts, b-tagging and by reconstructing 
the invariant masses of the two top squarks \cite{shibu}.

Another interesting  signal arises if one of the produced top squarks  
decay via an RPV mode (a) while the other decays via the loop induced mode Eq.
(22) followed by the LSP decay leading to
\\ 
\beq
 \tilde t_1   \tilde t_1^* \ar l_i^{+}  b  c \nu b \bar b ~~  or
~~ l_i^{-}  \bar b  \bar c \nu b \bar b
\eeq
 where i=1,2,3 as before  and $L_i$ denotes  the above
signal. The number of signal events for different $i$ are presented in 
Table [6] using the same inputs as in Table [5]. We have included 
leptons of both signs in the signal.    

\begin{table}[!htb]
\begin{center}

\begin{tabular}{|c|c|c|c|}
       \hline
          &$L_{1}$ & $L_{2}$ & $L_{3}$       \\

        \hline
 $\#$ of events (A) &152 &121  &804 \\
        \hline
 $\#$ of events (B) &206 &20  &789 \\
        \hline
 $\#$ of events (C) &198 & 829 & 35\\

        \hline
\end{tabular}
\end{center}
\caption{ Typical sizes of various signals from top squark pair production at 
Tevatron RUN II when one of them decays into a RPV channel while the other into
the loop induced mode followed by LSP decay.}
\end {table}

 The top squark mass can be reconstructed by
the invariant masses  of the two hardest leptons and jets \cite{shibu} in any
$L_{ij}$ type event. 
The upper edge 
of the invariant mass spectrum of the two jets with lowest and next to lowest
energy in any $L_i$ type signal may provide information about the LSP mass.

If the parameter $K_2$, which sets the scale of the one loop mass 
matrix, is decreased keeping the parameters of the gaugino sector fixed,
the allowed values of $\llamp_{i33}$ couplings increase and apparently larger
BRs of the RPV modes are allowed. However, we shall illustrate the 
constrained nature of the model in the next paragraph
and show that the above BRs  cannot be 
arbitrarily 
large inspite of many free RPC parameters  in the model. 

The parameter $K_2$ decreases for higher values of the common squark mass. 
In practice for a fixed set of gaugino parameters the common 
squark mass cannot be increased significantly without violating the top 
squark NLSP condition. Of 
course larger values of the  $A_t$ parameter may restore the NLSP 
condition. But larger values of $A_t$ tend to violate the CCB condition. 
Finally the parameter $M_A$ can be increased to satisfy the CCB condition
but as noted earlier that would enhance the loop decay width as well and 
the BRs of the RPV modes will still be suppressed. Thus once we know the 
parameters of the gaugino sector from complementary experiments 
and observe RPV decays of the top 
squark NLSP, the predicted BR of these modes cannot be made arbitrarily 
large by adjusting the common squark mass or $M_A$.

The BR of the RPV decays increase significantly if we consider model 3
with mixed $\tilde\chi_1^{\pm}$ and $\tilde\chi_2^0$. In this case the 
parameter C increases
substantially compared to it's typical magnitude in model 2  while 
the loop 
level mass matrix has a smaller $K_2$ due to a smaller higgsino mass 
parameter  $\mu$. Thus the loop 
level mass matrix can 
be effective only for larger values of the trilinear RPV couplings.  

We demonstrate these effects with the same parameter set as above except 
that we take $\mu$ = 210.0 GeV. We present the corrsponding  histogram 
in Fig. 2. It follows that the entire APS correspond to
 larger BRs of the  RPV modes ( including all leptonic channels). 

 For illustrating the signals at the LHC we consider a new parameter 
space leading  to a heavier top squark NLSP beyond the 
reach of the Tevatron. We have chosen $m_{\tilde t_1}$ = 350.65 GeV,
 $M_1=310$, $M_2=400$, $\mu=800$,
$\tan \beta =5$, $M_{\tilde q}=550 $, $M_{\tilde l} =450$,
$A_b=1000$, $A_{\tau}=1000$, $A_t=1350 $ and $M_{A}=500$ where all
masses and mass parameters are in GeV . At the LHC the top squark pair 
production cross section is 4.28 $pb$
for this $m_{\tilde t_1}$.

 We have chosen the following sets of RPV parameters 
consistent with the $\nu$ data which reflects
the same  characteristics as in A) - C) listed above.  \\

\noindent
A) $\llamp_{133}=2.92 \times 10^{-5}$,~$\llamp_{233}=2.27 
\times
10^{-5}$ and
$\llamp_{333}=6.27 \times 10^{-5}$ \\
B)   $\llamp_{133}=3.4 \times 10^{-5}$,
~$\llamp_{233}=6.6 \times 10^{-5}$ 
and
$\llamp_{333}=1.83 \times 10^{-5}$\\
C) $\llamp_{133}=3.39 \times 
10^{-5}$,
~$\llamp_{233}=1.92 \times 10^{-5}$ and$\llamp_{333}=6.48 \times 10^{-5}$\\
The corresponding BRs are shown in Table [7].

\begin{table}[!htb]
\begin{center}

\begin{tabular}{|c|c|c|c|c|c|}
       \hline
        Decay modes &$\tilde t_1  \ar  e^+   b$ & $\tilde t_1  \ar  \mu^+
 b$  & $\tilde t_1   \ar  \tau^+   b$  & loop & 4 body     \\
        \hline
BR(A)          &0.12  &0.075  &0.57  &0.23 &0.001 \\
BR(B)          &0.16  &0.59  &0.045  &0.21 &0.001 \\
BR(C)          &0.16  &0.05  &0.58  &0.21 &0.001 \\

        \hline
\end{tabular}
\end{center}
\caption{BRs of different decay modes of the lighter top squark in Model 2 
 ( see text for the choice of parameters ) }
\end{table}

Using the BRs in Table [7]
and a representative integrated luminosity of $30 fb^{-1}$ one can easily
calculate the number of various signal events. For example, we obtain 
44696 $L_{22}$ and 31817 $L_{2}$ events with the parameter set B.
 
\section*{5. LSP decay} 
As discussed in the introduction unless $m_{\LSP} > m_{t}$, LSP decay 
alone cannot provide detailed information about the underlying model of 
$m_{\nu}$. 

 Of course the decay  of the
LSP  into the channel in Eq. (1) may provide circumstantial 
evidences in favour of an underlying RPV model of 
neutrino mass. For example, if
$\LSP$ is assumed to be the LSP, then $\chplus_1 \chminus_1 $
and   $\chplus_1 \LSP $ production
followed by appropriate decay chains ending in LSP decays
are indicative of an underlying  model of neutrino mass \cite{barger}.
In Ref.~\cite{barger} the prospect of observing this signal at
RUN II was studied. It was concluded that
this signature can be probed up to $\mhf$ = 230~GeV(320 GeV) with an
integrated luminosity of 2 $fb^{-1}$ ( 30 $fb^{-1}$ ). Here $\mhf$
is the common gaugino mass at the GUT scale. However, the lepton number 
violating nature of the decay can not be established by the data.

It may be noted that the pure RPC decay  $\LSPsecond
\ar \LSP b \bar b $, which may have a large BR if one of the bottom
squark mass eigenstates happens to be lighter than the other squarks at 
large
$\tb$, has collider signatures very similar to the decay
of Eq.(\ref{lsp3bdk}). This is especially so if the LSP mass is much
smaller than $m_{\LSPsecond}$ which is quite common in models with 
non-universal gaugino masses. Thus one has to worry about the possibility
of RPC SUSY faking the RPV signal.

Using  model 1) (see section 4)  we have chosen the following 
of RPC parameters : \\
    $M_1=710$, $M_2=800$, $\mu=395$,
$\tan \beta =5$, $M_{\tilde q}=500 $, $M_{\tilde l} =450$,
$A_b=1000$, $A_{\tau}=1000$, $A_t=800 $ and $M_{A}=300$ where all
masses and mass parameters are in GeV. This leads to $\MST$ = 391 
GeV and $m_{\LSP}$=388 GeV 
 Since  $m_{\LSP} > m_{t}$ the  following 
additional decay modes open up :
\be
(a)~ \LSP \ra ~e ~\bar b ~t
\ee
\be
(b)~ \LSP \ra ~\mu ~\bar b ~t
\ee
\be
(c)~ \LSP \ra ~\tau ~\bar b ~t
\ee   
In Fig. 3 we present the BRs of the four competing decay modes of the LSP
and the corresponding number of allowed solutions out of $10^8$ sets of 
randomly generated RPV parameters.

From the sets allowed by the neutrino data we have chosen the following 
RPV parameters :
 $\llamp_{133}=5.43 \times 10^{-5}$,~$\llamp_{233}=10.19
\times 10^{-5}$ $\llamp_{333}=0.87 \times 10^{-5}$. The resulting BRs are
presented in column 2 of Table [8]. Here, the three neutrinos carry the 
$\met$.
                                                                                
\begin{table}[!htb]
\begin{center}
                                                                                
\begin{tabular}{|c|c|c|}
       \hline
        Decay modes & BR ($\%$)in Higgsino model &BR ($\%$)in Gaugino model\\ 
  
         \hline

$\LSP \ar \met b \bar b$       &12.5   &79.7 \\
$\LSP \ar t e \bar b$          &19.3   &3.7  \\
$\LSP \ar t \mu \bar b$        &67.7   &16.2 \\
$\LSP \ar t \tau \bar b $      &0.5     &0.4   \\
                                                                    
        \hline

\end{tabular}

\end{center}

   \caption{BRs of different decay modes of the LSP in higgsino
dominated model ( Model 1 ) and the gaugino dominated model (Model 2 ) ( see 
text for the choice of parameters).}

\end{table}

 We next compare this result
with the gaugino dominated scenario (Model 2).  We choose
$M_1=388$, $M_2=500$, $\mu=800 $ and $A_t= 870$ keeping all other 
RPC
parameters same as that in the  Higgsino type model. The choice of
parameters is dictated by the fact that the masses of the LSP and different
squarks remain practically the same in the two models being compared. The BRs in this
model are presented in the last column of Table [8] The BRs of the 
 modes involving charged leptons and the t quark are significantly 
enhanced in the Higgsino model because of 
the large top Yukawa coupling.

We next study  gaugino pair production followed by  cascade decays
 with RPC and RPV parameters as in model(1) quoted above. We consider 
the following signal \\
\beq
 G_{\alpha} G_{\beta}  \ar l_i^{(\mp)}  l_j^{(\mp)} \bar b \bar b t t  X
\eeq
where $G_{\alpha} G_{\beta}$ represents any pair of electroweak gauginos  
, i,j = 1,2,3 for e, $\mu$ and $\tau$ respectively,
  $L_{ij}$  represents the number of this signal and $X$
denotes any other particles produced.
The  number of signal events
 from different gaugino pair production at LHC are presented
in Table [9]. The production cross-sections involved are calculated 
using {\tt CalcHEP v2.1} \cite{CALCHEPcompc4}. 
$L_{12}$ includes e 
and $\mu$ events with all possible charge combinations  and $L_{11/22}$ 
represents events with same sign di-leptons only.

\begin{table}[!htb]
\begin{center}

\begin{tabular}{|c|c|c|c|c|}
       \hline
        Gaugino pair &$\sigma$($\pb$) & $L_{11}$ & $L_{22}$ &  
$L_{12}$  \\

        \hline
$\LSP$ $\LSPsecond$ &18.5 $\times 10^{-3}$ &206 &2551 &5808\\
       \hline
$\chonepl$ $\chonemi$&68.2 $\times 10^{-3}$ &762 &9405  &21416 \\
       \hline
$\chonemi$$\LSPsecond$&10.2 $\times 10^{-3}$ &113 &1406  &3200 \\
        \hline
$\chonepl$$ \LSPsecond$&23.4 $\times 10^{-3}$ &261 &3226 &7348\\
        \hline
$\LSP$ $\chonemi$&10.6 $\times 10^{-3}$ &118 &1461  &3328 \\
        \hline
$\LSP$ $\chonepl$&24.6 $\times 10^{-3}$ &274 &3392 &7724 \\

        \hline
\end{tabular}
\end{center}
   \caption{Number of $L_{11}$,$L_{22}$ and $L_{12}$ events (see text) 
arising from various gaugino pair production at LHC followed by LSP 
decay}
\end{table}

In Fig. 4 we present the LSP decay length vs the number of allowed 
solutions in Model 1. It is seen that apart from a small region of the APS 
the decay will be inside the detector.

\section*{6. Conclusions}

In this paper we have studied the APS of the RPV parameters 
in a model of neutrino mass subject to the
constraints imposed by the neutrino oscillation data \cite{global}.
The model we have considered has three bilinear RPV couplings $\mu_i$
and three trilinear couplings $\llamp_{i33}$, where $i$ is the lepton 
index,
and we work in a basis where the sneutrino vev is zero. As expected from 
the upper bounds on the $\llamp_{i33}$ couplings 
obtained by the earlier analyses 
\cite{abada}, we find for representative choice of the RPC parameters
, the allowed magnitudes of these couplings are indeed very small (see 
section 3). Our analyses reveal that even the currently available
$\nu$-oscillation data with relatively large errors are
 quite restrictive. Out of many randomly
generated sets of RPV parameters consistent with the upperbounds on them
\cite{abada},  only a few are allowed by the data.

Moreover, we also identify  six  generic scenarios, consistent with 
neutrino 
data, leading to distinctive collider signatures. These scenarios are
listed as $a_1)$ - $c_2)$ in section 3. In each scenario there is one 
small
bilinear parameter $\mu_i$ and a characteristic hierarchy of  $\llamp_{i33}$ 
couplings. Thus if a few 
direct RPV decay modes of any sparticle are observed their relative BRs would
reflect the  underlying model of 
neutrino mass. 
We have studied the 
decay modes of the top squark NLSP both at Tevatron RUN II and at the 
LHC
( section 4).
Over the entire APS the BRs of the RPV decays are found to be significant
for representative values of the RPC parameters
and the hierarchy among them can potentially reveal the underlying model 
of neutrino mass. The LSP decays can provide similar information if the
LSP mass is larger than $m_t$ (section 5).  

Hopefully RPV decays in the above channels will be seen at hadron colliders and 
at least some parameters of the RPC sector
can be measured kinematically ( 
some
 examples are discussed in section 4 ). The accurate measurement of the 
relative BRs
at the ILC will then provide an exciting program for probing the origin 
of neutrino mass.

In this paper we have chosen  a particular set of RPV parameters 
in the physical basis at the
 weak scale for explaining the current $\nu$- oscillation data. Following 
the standard practice we have further assumed that the magnitudes of all 
other RPV parameters are negligible. There is, however, another exciting 
possibility. Let a few relatively large RPV parameters not directly related 
to the
$\nu$ - sector, be generated at a high scale (say $M_{GUT}$) by any 
suitable
 mechanism ( for examples by effective operators \cite{Hall}
 ). The renormalization group
( RG ) evolution of these parameters down to the weak scale can then 
generate the
parameters $\mu_i$ and $\llamp_{i33}$ having magnitudes severely suppressed
compared to the input parameters \cite{ADAKJS}. This happens due to the 
flavour violation ( non-diagonal Yukawa couplings matrices ) 
inevitably present in the
 quark sector. It was shown in \cite{Dreiner,ADJPAK} that any three input 
couplings at $M_{GUT}$ with different lepton indices chosen from $\llamp_{i13}$
 and $\llamp_{i23}$, i=1,2,3, can induce 
a viable $\nu$- mass matrix at the weak scale. Thus at the weak scale there 
may be some relatively large $\llamp$ couplings in addition to the 
small parameters 
underlying $\nu$- physics. Moreover when the quark mass matrices are 
diagonalized at the weak scale rotations on the quark fields may further 
generate new RPV couplings \cite{ADJPAK} relevant for $\nu$- physics from the above 
relatively large $\llamp$- type couplings. Of Course such rotations may also 
induce other $\llamp$ -type couplings leading to unacceptable flavour changing
neutral currents \cite{Agashe}. Care should, therefore, 
be taken in choosing the 
input parameters at $M_{GUT}$. Because of the relatively large $\llamp$- type
couplings the RPV collider phenomenology and some rare decays of K-mesons 
and $\tau$-leptons would be rather spectacular in these 
models. Several specific models and their associated phenomenologies were 
discussed with numerical illustrations in \cite{ADJPAK}.

\vspace*{5mm}
{\bf Acknowledgment}:

SPD acknowledges support from the project (SP/S2/K-10/2001) of the 
Department of Science and Technology (DST), India. AD acknowledges 
support from the project (SR/S2/HEP-18/2003) of the DST, India.  
SP's work was supported 
by a fellowship from Council of Scientific and Industrial 
Research (CSIR), India.

\begin{figure}[!htb]
\vspace*{-4.0cm}
\hspace*{-3.0cm}
\mbox{\psfig{file=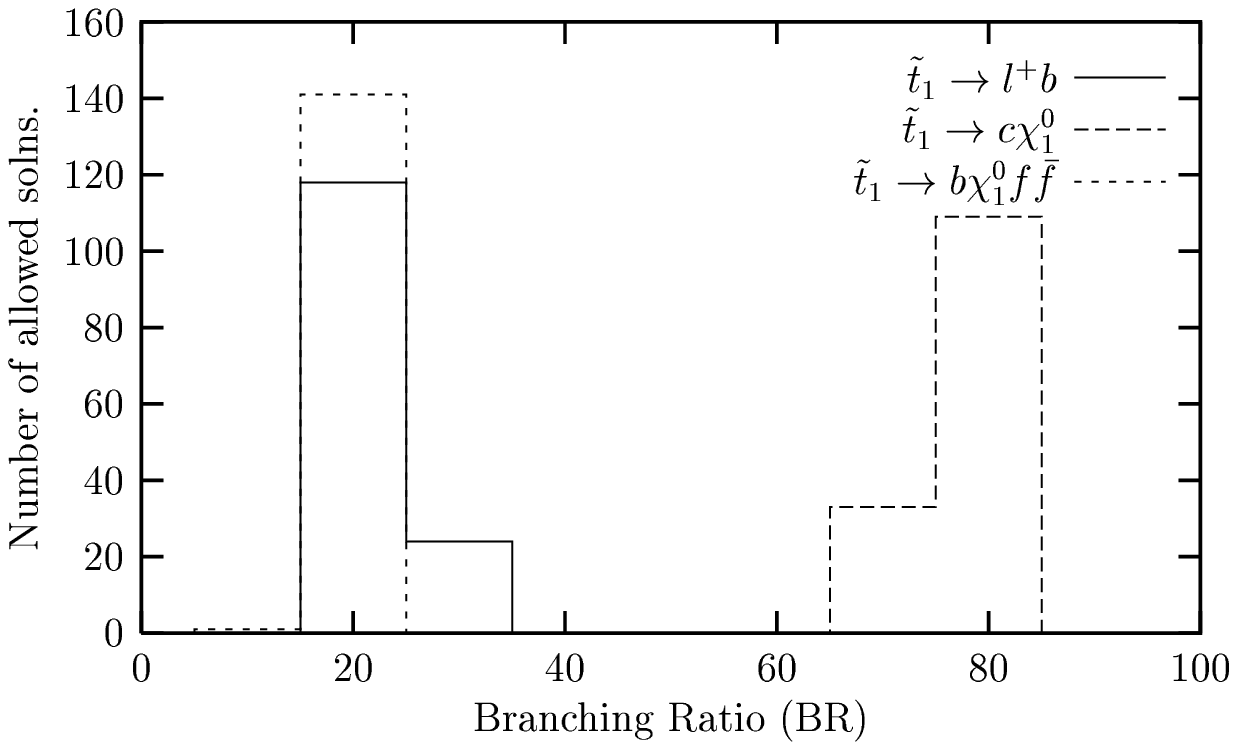,width=20cm}}
\vspace*{-16.2cm}
\caption{\small The branching ratios($\%$) of the three competing decay modes   vs 
  the number of allowed solutions in Model 2 ( see text for the parameters used
 ). }
\end{figure}

\begin{figure}[!htb]
\vspace*{-4.0cm}
\hspace*{-3.0cm}
\mbox{\psfig{file=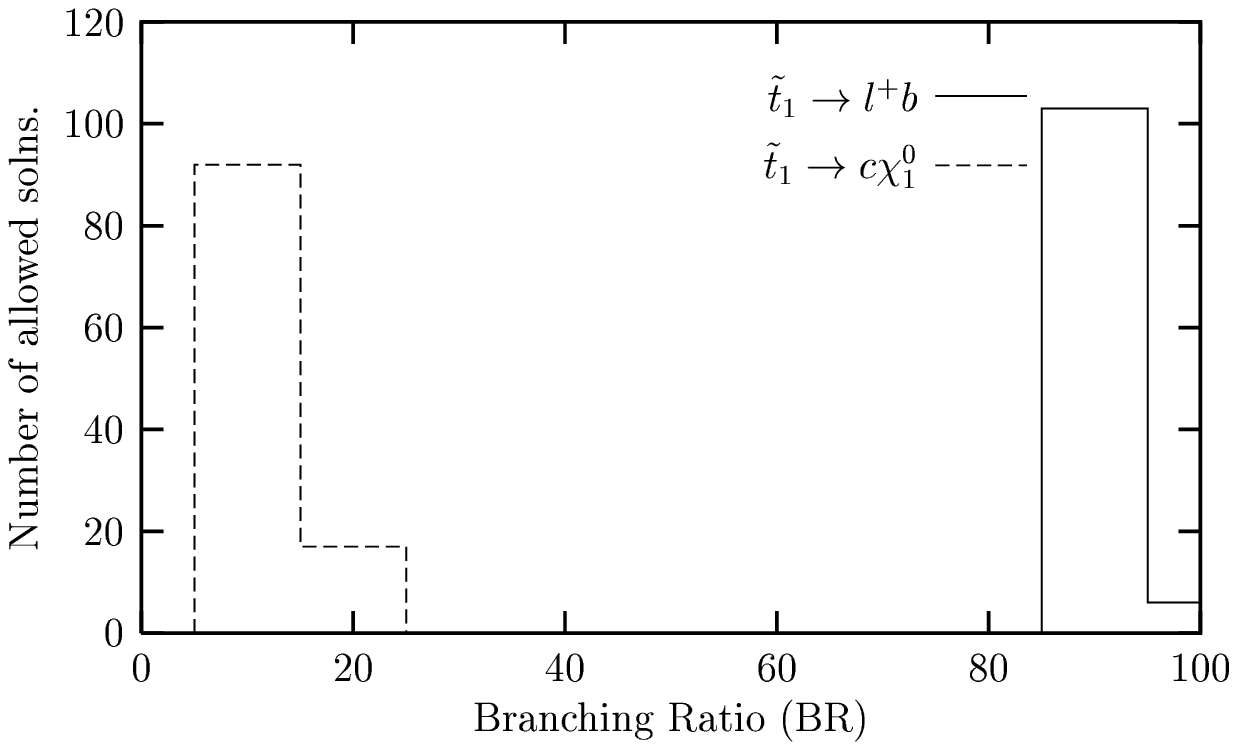,width=20cm}}
\vspace*{-16.2cm}
\caption{\small The branching ratios($\%$) of the two competing decay modes  vs 
  the number of allowed solutions in Model 3 ( see text for the parameters 
used ). }
\end{figure}
\begin{figure}[!htb]
\vspace*{-4.0cm}
\hspace*{-3.0cm}
\mbox{\psfig{file=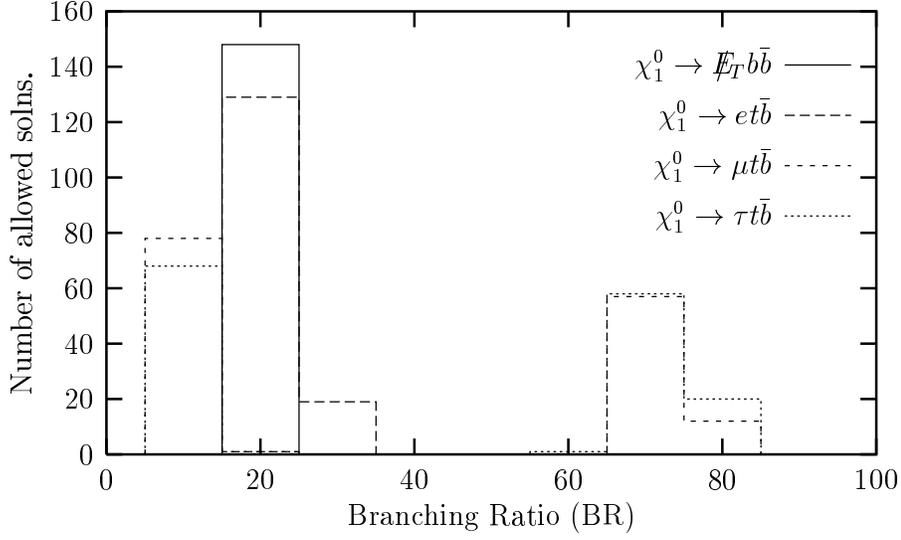,width=20cm}}
\vspace*{-16.2cm}
\caption{\small The branching ratios ($\%$) of competing LSP decay modes vs 
  the number of allowed solutions in Model 1 (see the text for the parameters
used ). }
\end{figure}

\begin{figure}[!htb]
\vspace*{-4.0cm}
\hspace*{-3.0cm}
\mbox{\psfig{file=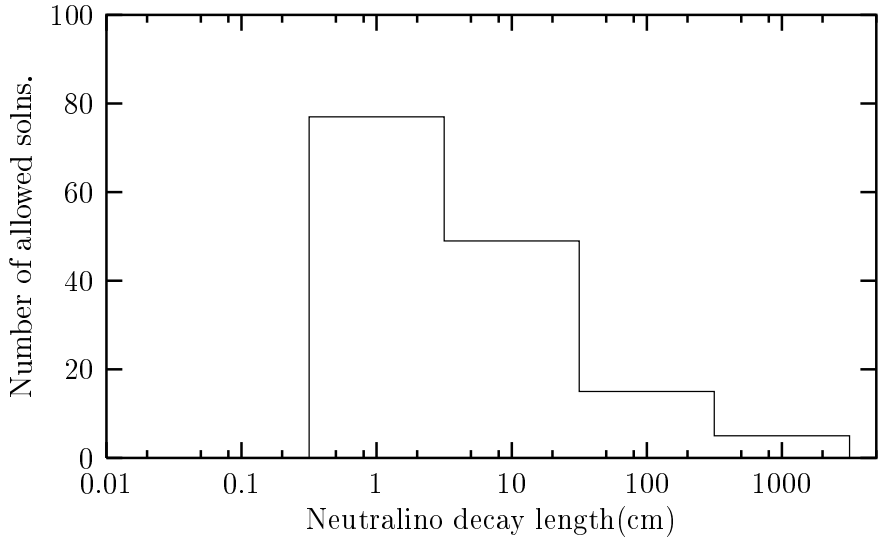,width=20cm}}
\vspace*{-16.2cm}
\caption{\small Neutralino decay length vs 
 the number of allowed solutions in Model 1 (see the text for the parameters
used ). }
\end{figure}

\end{document}
%%%%%%%%%%%%%%%%%%%%%%%%%%%%%%%%%%%%%%%%%
%% This is the order of the figures:
%fig2.ps
%fig3.ps
%fig4.ps
%fig5.ps